\documentclass[superscriptaddress, aip]{revtex4}
\usepackage[T1]{fontenc}
\usepackage{amsmath, amssymb, amstext}
\usepackage{graphicx}
\usepackage[utf8]{inputenc}
\usepackage{multirow}
\usepackage{physics} 
\usepackage{caption}
\usepackage[caption=false]{subfig} 
\usepackage{multirow}
\usepackage{array}
\renewcommand{\arraystretch}{2}

\def\be{\begin{equation}}
\def\ee{\end{equation}}
\def\bea{\begin{eqnarray}}
\def\eea{\end{eqnarray}}

\def\f{\frac}

\def\m{{\rm m}}

\makeatother

\begin{document}

\title{Qualitative dynamics of interacting vacuum cosmologies}

\author{Chakkrit Kaeonikhom}
\affiliation{Institute of Cosmology and Gravitation, University of Portsmouth,
Dennis Sciama Building, Portsmouth, PO1 3FX, United Kingdom}

\author{Phongsaphat Rangdee}
\affiliation{Department of Physics, School of Science, University of Phayao, Phayao, 56000, Thailand}

\author{Hooshyar Assadullahi}
\affiliation{Institute of Cosmology and Gravitation, University of Portsmouth,
Dennis Sciama Building, Portsmouth, PO1 3FX, United Kingdom}
\affiliation{School of Mathematics and Physics, University of Portsmouth, Lion Gate Building, Lion Terrace, Portsmouth, PO1 3HF, United Kingdom}

\author{Burin Gumjudpai }
\affiliation{The Institute for Fundamental Study ``The Tah Poe Academia Institute", Naresuan University, Phitsanulok 65000, Thailand}
\affiliation{Thailand Center of Excellence in Physics, \\ Ministry of Higher Education, Science, Research and Innovation, Bangkok 10400, Thailand}
\affiliation{Nakhon Sawan  Studiorum, Centre for Theoretical Physics \& Natural Philosophy, Mahidol University, Nakhon Sawan Campus,  Phayuha Khiri, Nakhon Sawan 60130, Thailand}

\author{Jascha A. Schewtschenko}
\affiliation{Institute of Cosmology and Gravitation, University of Portsmouth,
Dennis Sciama Building, Portsmouth, PO1 3FX, United Kingdom}

\author{David Wands}
\affiliation{Institute of Cosmology and Gravitation, University of Portsmouth,
Dennis Sciama Building, Portsmouth, PO1 3FX, United Kingdom}

\begin{abstract}
We present a phase-space analysis of the qualitative dynamics cosmologies where dark matter exchanges energy with the vacuum component. We find fixed points corresponding to power-law solutions where the different components remain a constant fraction of the total energy density and given an existence condition for any fixed points with nonvanishing energy transfer. For some interaction models we find novel fixed points in the presence of a third noninteracting fluid with constant equation of state, such as radiation, where the interacting vacuum+matter tracks the evolution of the third fluid, analogous to tracker solutions previously found for self-interacting scalar fields. We illustrate the phase-plane behavior, determining the equation of state and stability of the fixed points in the case of a simple linear interaction model, for interacting vacuum and dark matter, including the presence of noninteracting radiation. We give approximate solutions for the equation of state in matter- or vacuum-dominated solutions in the case of small interaction parameters.
\end{abstract}
\maketitle

\section{Introduction}
\label{Section:Intro}

The apparent late-time acceleration of our Universe \cite{Riess:1998cb,Perlmutter:1998np}, in the context of a homogeneous and isotropic cosmology governed by Einstein's equations, requires some form of dark energy with negative pressure. The simplest explanation would seem to be a positive energy density associated with empty space and hence undiluted by the cosmic expansion, i.e., vacuum energy 
with energy-momentum tensor $\check{T}_\mu^\nu=-V\delta_\mu^\nu$, 
density, $\rho_\mathrm{V}=V$ and pressure $P_\mathrm{V}=-V$. Together with the standard model of particle physics and an additional form of nonrelativistic (cold) dark matter, this forms the basis of the standard $\Lambda$CDM cosmology. However the remarkably small observed value of  the dark energy density today has led many authors to consider models of dark energy where the dark energy density evolves in time. Typically this is done by introducing additional dynamical degrees of freedom, e.g., quintessence \cite{Caldwell:1997ii,Peebles:2002gy,Copeland:2006wr}.

In this paper we consider the qualitative dynamics of cosmological models with time-dependent vacuum energy density, but without necessarily including any additional degrees of freedom other than the standard model particles plus cold dark matter. Instead we allow the vacuum to exchange energy with dark matter, leading to an alternative scenario with a time-dependent vacuum energy~\cite{Bertolami:1986bg,Freese:1986dd,Chen:1990jw,Carvalho:1991ut,Berman:1991zz,Wetterich:1994bg,Shapiro:2000dz,Sola:2005nh,Sola:2011qr,Barrow:2006hia,Wands:2012vg,DeSantiago:2012xh}. 
Such a time-dependent vacuum energy can arise from renormalization group running of the vacuum energy, which may also lead to running of Newton's gravitational constant~\cite{Shapiro:2000dz,Babic:2004ev,AgudeloRuiz:2019nnm}, although we do not consider such effects here. Similarly we do not include restrictions on the nature of the energy transfer in such models \cite{AgudeloRuiz:2020kox}.
The effective equation of state for the interacting vacuum and dark matter cosmology then depends on the form of this energy transfer, $Q=\dot{V}$. 

For example for vacuum energy interacting with cold dark matter with an energy transfer $Q=3\alpha H\rho_\mathrm{m}V/(\rho_\mathrm{m}+V)$, where $\rho_\mathrm{m}$ is the matter density, $H$ is the Hubble rate and $\alpha$ a dimensionless parameter, we recover the effective equation of state for the generalized Chaplygin gas (gCg) \cite{Kamenshchik:2001cp,Bento:2002ps}, but it originates from a nontrivial energy transfer between matter and vacuum energy \cite{Bento:2004uh}. While the two descriptions are equivalent at the level of the homogeneous background they may differ at the level of inhomogeneous perturbations, and specifically the effective speed of sound \cite{Wands:2012vg,Carneiro:2014uua,Borges:2017jvi}. 
In the matter-dominated limit, the gCg interaction model reduces to $Q\propto HV$ \cite{Salvatelli:2014zta,Martinelli:2019dau,Hogg:2020rdp} and reduces to $Q\propto H\rho_\mathrm{m}$ \cite{Shapiro:2003ui,EspanaBonet:2003vk,Wang:2004cp,Alcaniz:2005dg,Sola:2017jbl} when the vacuum energy is dominant. This leads us to consider a simplified interaction model $Q=\alpha H\rho_\mathrm{m}+\beta HV$, with two dimensionless parameters $\alpha$ and $\beta$, which we will use in this paper to illustrate the qualitative effects of an interaction at early and late times. 
Similar models have previously been studied in the context of interacting dark energy models~\cite{Sadjadi:2006qp, CalderaCabral:2008bx, Quartin:2008px, Quercellini:2008vh}.
However we are also able to consider some of the qualitative properties of models solely in terms of the interaction at fixed points in the phase space, applicable to more general interaction models.

A phase-space analysis enables us to study fixed points of the evolution of interacting vacuum and dark matter cosmologies, with a spatially flat 
Friedmann-Lemaitre-Robertson-Walker
(FLRW) metric with scale factor, $a$. We use dimensionless variables~\cite{Copeland:1997et,Billyard:2000bh,Heard:2002dr,Bahamonde:2017ize} for the fluid and vacuum energy densities relative to the critical density (determined by the Hubble expansion, $H=\dot{a}/a$). Fixed points in this phase space correspond to scale-invariant (``scaling'') solutions and, in the limiting case of a constant Hubble rate, de Sitter. We give analytic (power-law) solutions for the scale-invariant fixed points and study their stability.

The layout of the paper is as follows. In Sec.~\ref{Section:DynamicalVariables} we define our dimensionless variables and give general conditions for fixed points in the presence of an interaction. In Sec.~\ref{sec:ref2fluid} we characterize fixed points in the presence of two interacting fluids. In this case the Friedmann constraint reduces the phase space to a one-dimensional space. In Sec.~\ref{sec:ref3fluid} we consider three-fluid cosmologies including a noninteracting barotropic fluid which leads to the presence of additional fixed points in a two-dimensional phase-space. We conclude in Sec.~\ref{sec:discuss}, illustrating one of the regimes in our parameter space with a late-time accelerating matter+vacuum cosmology emerging from a conventional early radiation-dominated era.

\section{Dynamical Variables}
\label{Section:DynamicalVariables}

Let us consider pressureless matter density, $\rho_m$, interacting with vacuum energy, $V$. The coupled energy conservation equations are then
%
\begin{align}
\dot{\rho}_\mathrm{V} &= +Q, \label{Eqn:ContinuityV} \\
\dot{\rho}_\mathrm{m} + 3H\rho_\mathrm{m} &= -Q \,.
\label{Eqn:Continuitym} 
\end{align}

We define dimensionless variables describing the matter and vacuum energy densities relative to the expansion
\begin{equation}
x \equiv \frac{\kappa\sqrt{\rho_\m}}{\sqrt{3}H} ,
\quad 
y \equiv \epsilon \frac{\kappa\sqrt{\epsilon V}}{\sqrt{3}H}\label{Var:xy}
\end{equation}
where we use $\kappa^2 = 8\pi G$, and 
\begin{equation}
	\epsilon=\left\{\begin{array}{ccc}{+1} & \text{for} & {V>0} \\ 
	{-1} & \text{for} & {V<0.}\end{array}\right.
\end{equation}
We allow for the possibility that the vacuum density may be positive or negative, corresponding to the choice of upper or lower signs throughout, while we will assume that the matter density remains non-negative throughout (otherwise we may have a negative number of particles or particles with a negative energy, leading to ghost-like instabilities~\cite{Carroll:2003st}).

We will study their dynamical evolution with respect the logarithmic expansion. In the following a prime denotes a derivative with respect to the logarithm of the scale factor:
\be
' \equiv \frac{d}{d\ln a} = \frac1H \frac{d}{dt} \,.
\ee
Thus an expanding universe ($H>0$) with positive vacuum energy ($V>0$) corresponds to the upper-right quadrant ($x>0$, $y>0$) and one with negative vacuum energy corresponds to the lower-right quadrant ($x>0$, $y<0$). In the following we will restrict our discussion to expanding universes ($x>0$).
As we are using the logarithmic scale factor as our time coordinate we may treat collapsing universes ($H<0$) simply as the time reverse of the expanding case. Note that a spatially flat cosmology with positive matter density can only have a turning point ($H=0$) in the presence of another component with negative energy density, corresponding to $y<0$ in this case.

The evolution equations can then be written as
\bea
\label{Eqn:Xprime-wq}
x' &=& \left( \frac32 w - \frac12 \frac{q}{x^2} \right) x \\
\label{Eqn:Yprime-wq}
y' &=& \left[ \frac32 (1+w) +\epsilon \frac12 \frac{q}{y^2} \right] y \,.
\eea
where we have defined the dimensionless energy transfer
\begin{equation}
q \equiv \frac{\kappa^2 Q}{3H^3}\,.
\label{def:q}
\end{equation}
The evolution of the Hubble rate is given by
\be
\label{Eqn:Hprime-w}
\frac{H'}{H} = -\frac32 (1+w) \,,
\ee
in terms of the effective overall equation of state
\begin{equation}
\label{def:w}
w \equiv \frac{P}{\rho} \,.
\end{equation}

Our choice of variables mimics those commonly chosen to study scaling solutions in scalar field cosmologies~\cite{Copeland:1997et,Billyard:2000bh,Heard:2002dr}. 
$x^2$ can be identified with the usual dimensionless density parameter for matter $\Omega_\mathrm{m}=x^2$ while the dimensionless vacuum energy parameter is $\Omega_\mathrm{V}=\epsilon y^2$.
Fixed points in the phase space thus correspond to scaling solutions with constant density parameters, $\Omega_\mathrm{m}$ and $\Omega_\mathrm{V}$, and constant overall equation of state $w$, which in turn implies, from Eq.~
\eqref{Eqn:Hprime-w}, a power-law solution for the scale factor
\begin{equation}
a\propto t^{2/3(1+w)} \,.
\label{Eqn:apropt}
\end{equation}

In the case of a noninteracting vacuum with $q=0$, we can immediately identify two familiar fixed points for $x$ and $y$. We have a fixed point at $x_p=1$, $y_p=0$ with $w=0$, i.e., a matter-dominated expansion ($\Omega_\mathrm{m}=1$) where $a\propto t^{2/3}$. We also have a vacuum-dominated  ($\Omega_\mathrm{V}=1$) fixed point at $x_p=0$, $y_p=1$ with $w=-1$, i.e., de Sitter expansion where $H'=0$. These will remain fixed points, i.e., asymptotic limits of the behavior in more general models allowing for $q\neq0$. 

More generally, we see from Eqs.~(\ref{Eqn:Xprime-wq}) and~(\ref{Eqn:Yprime-wq}) that an interacting fixed-point solution ($x'\rightarrow0$ and $y'\rightarrow0$) exists for nonzero $x\to x_p$ and $y\to y_p$ when
\begin{equation}
q = 3wx_p^2 = - 3(1+w)\epsilon y_p^2 
\label{Eqn:qnovel}
\end{equation}
%

To be able to close the dynamical phase space  $\{x,y\}$ we need to determine the overall equation of state, $w$ in Eq.~(\ref{def:w}), and dimensionless energy transfer, $q$ in Eq.~(\ref{def:q}), in terms of the dynamical variables. 
We shall therefore consider models in which $w$ and $q$ can then be expressed solely in terms of $x$ and $y$.
Many kinds of interaction models have been reviewed in Ref.  \cite{Bahamonde:2017ize}; however in this paper we will focus our attention on a simple linear interaction model
\begin{equation}
Q = \alpha H \rho_\mathrm{m} + \beta H V,
\label{modelQ}
\end{equation}
where $\alpha$ and $\beta$ are dimensionless coupling parameters. In terms of our dimensionless energy transfer \eqref{def:q} this corresponds to
\begin{equation}
q = \alpha x^2 +\epsilon \beta y^2.
\label{Eqn:qsimple}
\end{equation}

In addition we have not yet imposed the Friedmann constraint equation which determines the Hubble expansion in terms of the total energy density. This constraint therefore depends on the number of fluids contributing to the energy density. 
In what follows, we will consider the simple interaction model (\ref{modelQ}) in both a two-fluid system (interacting vacuum+matter) and a three-fluid system (interacting vacuum+matter, plus a third, noninteracting barotropic fluid).


\section{Two-fluid system}
\label{sec:ref2fluid}

In this section we will consider a spatially flat FLRW cosmology with pressureless dark matter interacting with the vacuum energy. The Friedmann equation
\begin{equation}
H^2 = \frac{\kappa^2}{3} \left( \rho_\mathrm{m} + V \right) \,,
\end{equation}
which thus constrains the variables $x$ and $y$ to lie on a one-dimensional curve
%
\begin{equation}
x^2 + \epsilon y^2 = 1.
\label{Eqn:ConstraintXY2fluid}
\end{equation}
The overall equation of state (\ref{def:w}) for the two-fluid case becomes
\begin{equation}
	w = \frac{-V}{\rho_\mathrm{m} + V } 
	= -\epsilon y^2.
	\label{Eqn:wx}
\end{equation}
The dimensionless deceleration parameter, $q_{\rm dec}\equiv -a\ddot{a}/\dot{a}^2=(1+3w)/2$, is thus given by
\begin{equation}
q_{\rm dec} = \frac{x^2}{2}-\epsilon y^2 \,.
\end{equation}

\subsection{Fixed points}

A fixed point, $x'\rightarrow 0$ and $y'\to0$ where $x\to x_p$ and $y\to y_p$, corresponds to a constant overall equation of state $w\to w_p$ given by \eqref{Eqn:wx} and 
we obtain a power-law solution given by \eqref{Eqn:apropt} for the cosmological scale factor as a function of cosmic time
\begin{equation}
a \propto t^{2/3x_p^2} \,.
\end{equation}

Substituting Eq.~\eqref{Eqn:wx} for $w$ into the condition for a novel fixed point \eqref{Eqn:qnovel} we obtain a simple existence condition for a two-fluid fixed point
\begin{equation}
q = -3 \epsilon x_p^2 y_p^2 \,
\label{Eqn:q2fluidcondition}
\end{equation}
for any interaction model $q$. We recover the familiar fixed points for a matter-dominated solution ($x_p=1$, $y_p=0$) and de Sitter ($x_p=0$, $y_p=1$) as fixed points where these coincide with zero interaction, $q\to0$. There may be additional fixed points for nonzero interaction, $q\neq0$, dependent on the interaction model.


\subsection{Simple interaction model}

The continuity equations for matter density and vacuum energy with the interaction model \eqref{modelQ} are given by
\begin{align}
\dot{\rho}_\mathrm{V} &= \alpha H\rho_\mathrm{m} + \beta H V, 
\label{Eqn:ContinuityVac} \\
\dot{\rho}_\mathrm{m} &= - (3 + \alpha)H\rho_\mathrm{m} -\beta HV
\label{Eqn:ContinuityMat}
\end{align}
In terms of our dimensionless variables we have
\begin{align}
x' &= \qty[- \frac{3 + \alpha}{2} - \epsilon \frac{\beta}{2}\frac{y^2}{x^2} + \frac{3}{2}x^2]x , \label{Eqn:Xprime} \\
y' &= \qty[\frac{\beta}{2}+\epsilon\f{\alpha}{2}\frac{x^2}{y^2} + \frac{3}{2}x^2 ]y \,.
\label{Eqn:Yprime}
\end{align}

We will consider separately the fixed points and their properties in simple cases such as $\alpha=0$ or $\beta=0$, and then the general case where $\alpha,\beta\neq0$. The fixed points are summarized for each case in Table \ref{tab:2-fluid} and we will analyze in each case separately in the following subsections.

\subsubsection{Case I: $\alpha \neq$ 0 and $\beta$ = 0}
\label{SubSection:CaseI}

The fixed-point existence condition \eqref{Eqn:q2fluidcondition} for this case is given by
\begin{equation}
\alpha x_p^2 = -3\epsilon x_p^2 y_p^2 \,.
\label{fixedpoint-case-I}
\end{equation}
Using the two-fluid constraint \eqref{Eqn:ConstraintXY2fluid} yields the fixed-point solutions
\begin{align}
\mathbf{Ia:}\quad\quad	x_{\mathrm{Ia}} &= 0, \quad y_{\mathrm{Ia}} = 1, 
\label{eq:xp-Ia} 
\\
\mathbf{Ib:}\quad\quad	x_{\mathrm{Ib}} &= \sqrt{1 + \frac{\alpha}{3}}, \quad y_{\mathrm{Ib}} = \epsilon\sqrt{-\epsilon\frac{\alpha}{3}}.
\label{eq:xp-Ib}
\end{align}
The fixed point \textbf{Ia} exists for all values of $\alpha$ while the point \textbf{Ib} exists for $\alpha>-3$. (The two points coincide for $\alpha=-3$.)

We determine their stability by introducing small perturbations $u$ and $v$ about the fixed points
\begin{equation}
x \rightarrow x_p + u, \quad y \rightarrow y_p + v
\end{equation}
and considering the evolution Eqs. \eqref{Eqn:Xprime} and \eqref{Eqn:Yprime} to first order in $u$ and $v$.
In addition, considering the two-fluid Friedmann constraint \eqref{Eqn:ConstraintXY2fluid} yields 
\begin{equation}
x_pu = - \epsilon y_p v.
\label{uv-constraint}
\end{equation}

At fixed point \textbf{Ia}, Eq.~\eqref{uv-constraint} shows that $x_{\mathrm{Ia}}=0$ implies $v=0$ at first order. Hence 
Eq.~\eqref{Eqn:Xprime} becomes
\begin{equation}
u' = \qty( -\frac{3 + \alpha}{2})u
\end{equation}
Thus $u\propto e^{\lambda_{\mathrm{Ia}}N}$ where the eigenvalue $\lambda_{\mathrm{Ia}}$ is given by
\begin{equation}
\lambda_{\mathrm{Ia}} = -\frac{3 + \alpha}{2}.
\label{eq:lambda-Ia}
\end{equation}
The point \textbf{Ia} is therefore a stable node ($\lambda_\mathrm{Ia}<0$)  if $\alpha>-3$ and an unstable node ($\lambda_\mathrm{Ia}>0$) for $\alpha<-3$.

The fixed point \textbf{Ib} exists for $\alpha>-3$. For $-3<\alpha<0$ it corresponds to positive vacuum energy $y_{\mathrm{Ib}}>0$, and for $\alpha>0$ it corresponds to negative vacuum energy $y_{\mathrm{Ib}}<0$. Considering a small perturbation $x\rightarrow x_{\mathrm{Ib}}+u = \sqrt{1+(\alpha/3)}+u$ and substituting this into Eq.~\eqref{Eqn:Xprime} for the point \textbf{Ib} at first order we obtain
\begin{equation}
 u' = (3+\alpha) u,
\end{equation}
and hence the eigenvalue is given by 
\begin{equation}
\lambda_\mathrm{Ib} = 3 + \alpha.
\label{eq:lambda-Ib}
\end{equation}
Given that this fixed point requires $\alpha>-3$ to exist, we see that it always corresponds to an unstable node, $\lambda_\mathrm{Ib}>0$.

\begin{figure*}
\begin{center}
\label{fig:caseI}
\includegraphics [width=\textwidth]{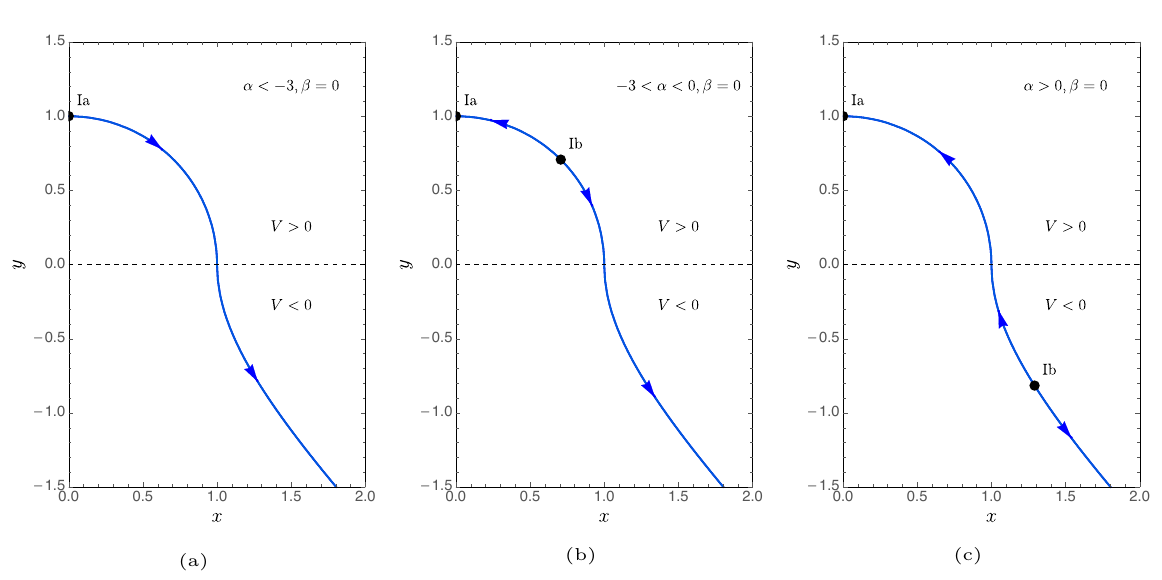}
\caption{\label{fig:2-fluid-beta0}The phase-space evolution of $(x,y)$ with respect to $N$ for the two-fluid system in case I ($\beta=0$), illustrating the existence and stability of the fixed points in each range of $\alpha$.}
\end{center}
\end{figure*}

The phase-space trajectories in this case are shown in Fig. \ref{fig:2-fluid-beta0}.
\begin{itemize}
\item
For $\alpha<-3$ the only fixed point is \textbf{Ia}; general solutions start at the vacuum-dominated fixed point \textbf{Ia} with $w_\mathrm{Ia}=-1$ but evolve towards $y<0$.
\item
For $\alpha>-3$ solutions start from an initial scaling solution \textbf{Ib} with overall equation of state \eqref{Eqn:wx} corresponding to $w_\mathrm{Ib}=\alpha/3>-1$ (with negative vacuum energy, $y_\mathrm{Ib}<0$ and $w_\mathrm{Ib}>0$ when $\alpha>0$, or positive vacuum energy and $y_\mathrm{Ib}>0$ and $w_\mathrm{Ib}<0$ when $\alpha<0$) and evolve towards either the late-time vacuum-dominated fixed point, \textbf{Ia} with $w_\mathrm{Ia}=-1$, or towards $y\to-\infty$, as shown in Fig. \ref{fig:2-fluid-beta0}.
\end{itemize}

%
\subsubsection{Case II: $\alpha$ = 0 and $\beta \neq$ 0}
\label{SubSection:CaseII}

In this case, the fixed point existence condition \eqref{Eqn:q2fluidcondition} is given by
\begin{equation}
\epsilon \beta y_p^2 = -3\epsilon x_p^2y_p^2 \,.
\end{equation}
Solving the above equation together with the constraint \eqref{Eqn:ConstraintXY2fluid}, the fixed points can be shown to be
\begin{align}
\mathbf{IIa:}\quad\quad	x_{\mathrm{IIa}} &= 1, \quad y_{\mathrm{IIa}} = 0, 
\label{eq:xp-IIa} \\
\mathbf{IIb:}\quad\quad	
x_{\mathrm{IIb}} &= \sqrt{-\frac{\beta}{3}}, \quad y_{\mathrm{IIa}} = 
	\epsilon\sqrt{\epsilon\left(1 + \frac{\beta}{3}\right)}.
	\label{eq:xp-IIb}
\end{align}

\begin{figure}[!h]
\centering
\includegraphics[width=\textwidth]{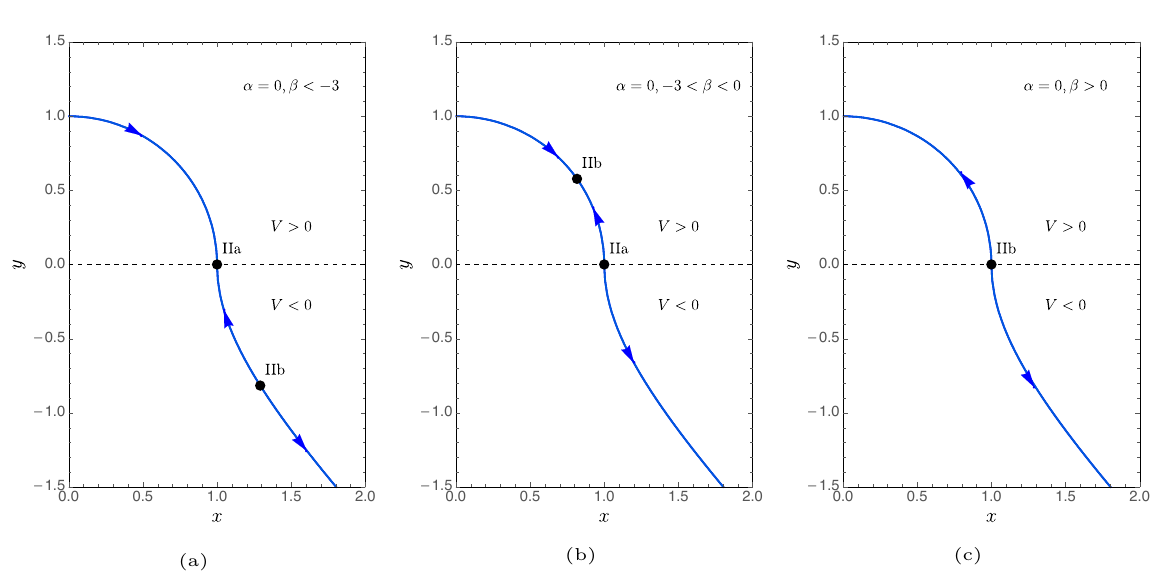}
\caption{\label{fig:2-fluid-alpha0}The phase-space evolution of $(x,y)$ with respect to $N$ for the two-fluid system in case II ($\alpha=0$), illustrating the existence and stability of the fixed points in each range of $\beta$.}
\end{figure}

The point \textbf{IIa} exists for all values of $\beta$ and corresponds to a matter-dominated solution with $w_{\mathrm{IIa}}=0$. However, the point \textbf{IIb} exists only for negative $\beta$;  it lies in the upper-right quadrant ($y>0$) with $V>0$ for $-3<\beta<0$ but lies in the lower-right quadrant ($y<0$) with $V<0$ for $\beta<-3$, as illustrated in Fig. \ref{fig:2-fluid-alpha0}.

We perform a similar stability analysis as the previous case I, but using the small perturbation $y \rightarrow y_p + v$ in this case. Near the fixed point \textbf{IIa} $(y_{\mathrm{IIa}}=0)$, Eq.~\eqref{uv-constraint} implies $u=0$ at first order, and Eq.~\eqref{Eqn:Yprime} to first order yields 
\begin{equation}
v' = \qty(\frac{3+\beta}{2})v,
\end{equation}
and hence $v\propto e^{\lambda_{\mathrm{IIa}}N}$ with the corresponding eigenvalue
\begin{equation}
\lambda_{\mathrm{IIa}} = \frac{3+\beta}{2}
\label{eq:lambda-IIa}
\end{equation}
implying the point \textbf{IIa} is stable for $\beta<-3$ and unstable for $\beta>-3$.

The other fixed point for $\beta<0$ is \textbf{IIb}.
The stability analysis in this case yields the eigenvalue
\begin{equation}
\lambda_\mathrm{IIb} = -(3+\beta)\,,
\label{eq:lambda-IIb}
\end{equation}
Thus the point \textbf{IIb} is stable for $-3<\beta<0$, corresponding to $y_\mathrm{IIb}>0$ and hence $V>0$ at the fixed point and $w_\mathrm{IIb}<0$, and it is unstable if $\beta<-3$, corresponding to $y_\mathrm{IIb}<0$, and $V<0$ and $w_\mathrm{IIb}>0$.

The phase-space trajectories in this case are shown in Fig. \ref{fig:2-fluid-alpha0}.
For $\beta>0$ the only fixed point is \textbf{IIa}; general solutions start at the matter-dominated fixed point \textbf{IIa} with $w_\mathrm{IIa}=0$ and evolve either towards $y\to-\infty$ for $V<0$, or would evolve to reach $x=0$ for $V>0$ after which the density would then become negative, which we consider to be unphysical.
When $-3<\beta<0$, solutions start from an initial matter-dominated state, \textbf{IIa}, and either end up at the scaling solution \textbf{IIb} with overall equation of state \eqref{Eqn:wx} corresponding to $w_\mathrm{IIb}=-(\beta/3)-1>-1$ or evolve towards $y\to-\infty$. 
For $\beta<-3$ the only physical initial state is the scaling solution \textbf{IIb} with $V<0$ and $w_\mathrm{IIb}=-(\beta/3)-1>0$, which is unstable and general solutions evolve either towards the matter-dominated solution \textbf{IIa} with $w_\mathrm{IIa}=0$ or towards $y\to-\infty$, as shown in Fig. \ref{fig:2-fluid-alpha0}.

\subsubsection{Case III: $\alpha, \beta \neq$ 0}

\label{SubSection:CaseIII}

\begin{figure}
\centering
	\includegraphics[width=0.5\textwidth]{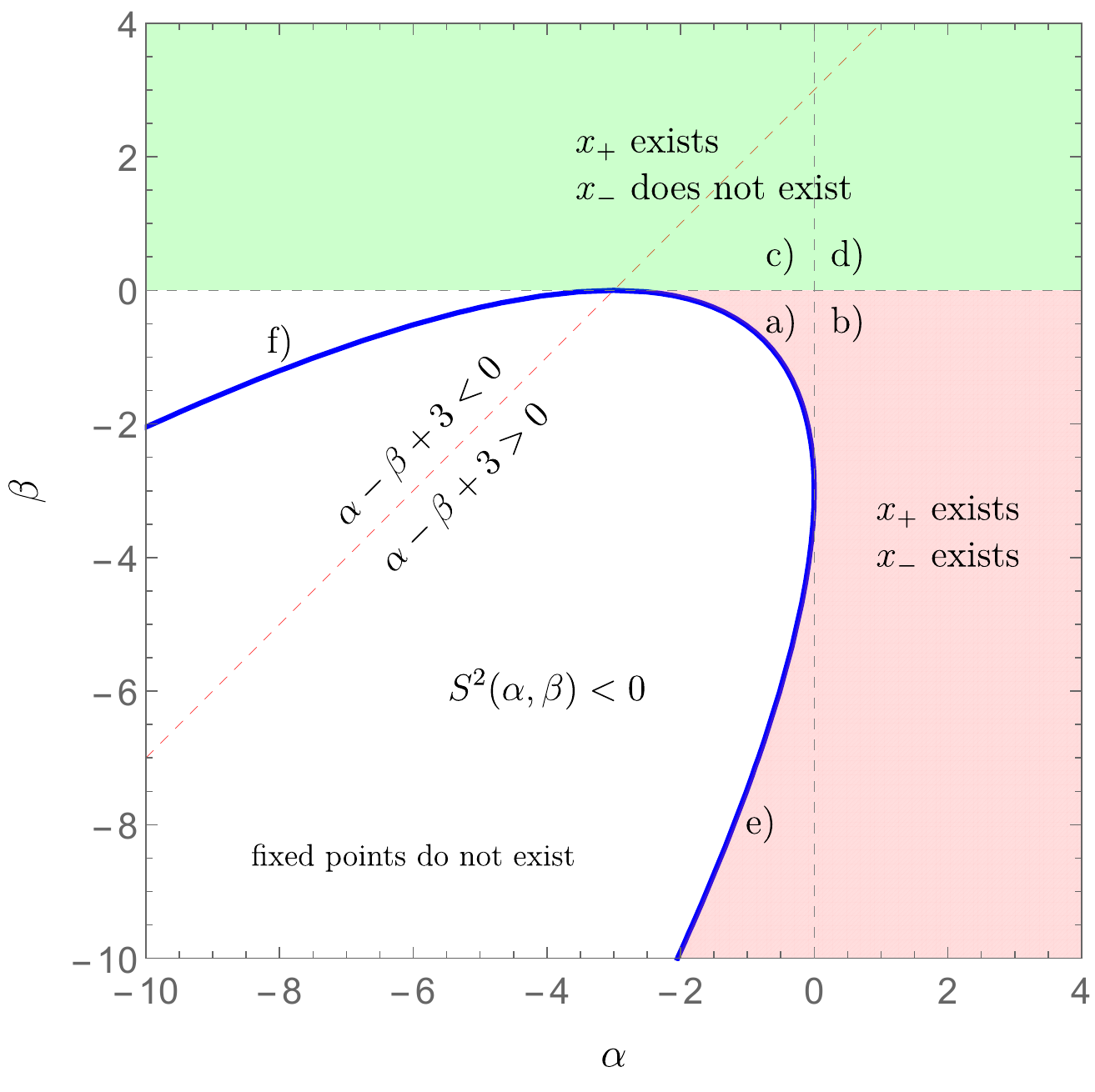}
\caption{\label{Fig:gen.case-condition}Different parameter regimes for the simple interaction model~\eqref{Eqn:qsimple} with both $\alpha\neq0$ and $\beta\neq0$. The solid blue curve denotes $S^2(\alpha,\beta)=0$ where $S(\alpha,\beta)$ is defined in Eq.~\eqref{Eqn:definedS}. The shaded regions denote the existence of both fixed points \textbf{III$\pm$} (red), only one fixed point \textbf{III+} (green), or no fixed points (white). The letters a)-f) denote the different parameter regimes illustrated in Fig.~\ref{Fig:gen.case-trajectories}.}
\end{figure}

In the case of nonzero $\alpha$ and $\beta$ for the simple interaction model~\eqref{Eqn:qsimple}, the existence condition \eqref{Eqn:q2fluidcondition} yields
\begin{equation}
\alpha x_p^2+\beta\epsilon y_p^2= -3\epsilon x_p^2y_p^2.
\end{equation} 
Applying the constraint \eqref{Eqn:ConstraintXY2fluid} then gives two possible fixed points
\begin{align}
\mathbf{III\pm:}\quad\quad
	x_{\mathrm{III}\pm} &= \sqrt{\frac{(\alpha - \beta + 3) 
	\pm S(\alpha,\beta)}{6}},\quad
	\label{xIII:defined}
	y_{\mathrm{III}\pm} = \epsilon\sqrt{\frac{(\beta-\alpha+3) 
	\mp S(\alpha,\beta)}{6\epsilon}} 
\,,
\end{align}
where we have defined
\begin{equation}
	S(\alpha,\beta)\equiv\sqrt{(\alpha+\beta + 3)^2 - 4\alpha\beta}
	\label{Eqn:definedS}
\end{equation}
and we require $S^2(\alpha,\beta)\geq0$.
$y_{\mathrm{III}\pm}$ is always real since we can choose the sign of $\epsilon$ in order to ensure $y_{\mathrm{III}\pm}$ is real. Therefore the fixed points exist if $S(\alpha,\beta)$ is real and if $x_{\mathrm{III}\pm}$ is real, which requires
\begin{equation}
	(\alpha-\beta+3) \pm S(\alpha,\beta)
	 \geq 0 \,.
	\label{Eqn:xpcondition}
\end{equation}
%
The different parameter regimes for the existence of the two fixed points are illustrated in Fig.~\ref{Fig:gen.case-condition}.
We find three regimes:
\begin{itemize}
\item
\textbf{III+} exists and \textbf{III-} does not exist for $\beta>0$.
\item
Both \textbf{III+} and \textbf{III-} exist if $S^2(\alpha,\beta)>0$, $\beta<0$ and $\alpha>\beta-3$.
\item
Neither \textbf{III-} nor \textbf{III+} exist if either $S^2(\alpha,\beta)\leq0$ or if $\beta<0$ and $\alpha<\beta-3$
\end{itemize}

%

Introducing small perturbations about the fixed points, $x\rightarrow x_{\mathrm{III}\pm}+u$ and $y\rightarrow y_{\mathrm{III}\pm}+v$, and writing Eqs. \eqref{Eqn:Xprime} and \eqref{Eqn:Yprime} to first order in $u$ and $v$, we have
\begin{equation}
\left(
\begin{array}{c} u' \\ v' \end{array}
\right) 
= 
\pm S(\alpha,\beta)
\left(
\begin{array}{c} u \\ v \end{array}
\right) 
\end{equation}
%
Thus the eigenvalues for the evolution of perturbations about each fixed point have the simple form
\begin{equation}
\lambda_{\mathrm{III}\pm} = \pm S(\alpha,\beta). 
\label{EigenValuesS}
\end{equation}
%
We find that \textbf{III+} is an unstable node with the eigenvalue $+S(\alpha,\beta)>0$, while \textbf{III-} is a stable node with the eigenvalue $-S(\alpha,\beta)<0$.

The equation of state at the fixed points is given by Eq. \eqref{Eqn:wx}, and hence
\begin{equation}
w_{\mathrm{III}\pm} = -\frac{3-\alpha+\beta \mp  S(\alpha,\beta)}{6} \,.
\label{wIII:defined}
\end{equation}

\begin{figure}[!h]
\centering
\includegraphics[width=\textwidth]{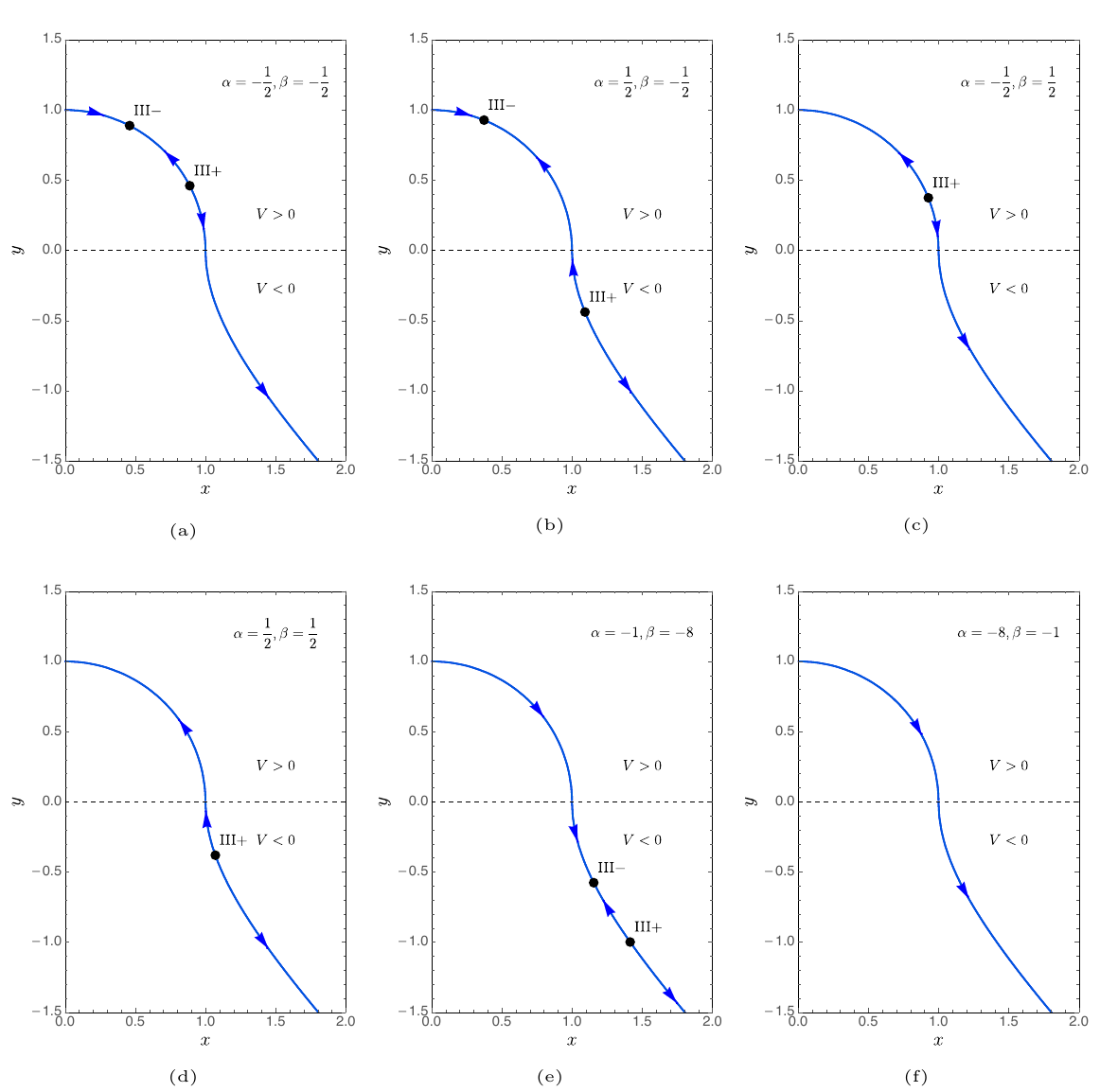}
\caption{The phase-space evolution of $(x,y)$ with respect to $N$ for the two-fluid system in case III ($\alpha\neq0$ and $\beta\neq0$), illustrating the existence and stability of the fixed points in different parameter regimes. \label{Fig:gen.case-trajectories}}
\end{figure}

The different possible trajectories in different parameter regimes are illustrated in Fig.~\ref{Fig:gen.case-trajectories}.
In parameter regime (a), for example, for small positive values of $\alpha$ ($\alpha\ll1$) and small negative values of $\beta$ ($-\beta\ll1$) solutions may start close to a matter-dominated scaling solution, $w_{\mathrm{III+}}\approx0$ and evolve to a vacuum-dominated scaling solution with $w_{\mathrm{III-}}\approx-1$.

%

\begin{table}[!h]
\caption{The fixed points, $x_p$ and $y_p$, and the eigenvalues for the two-fluid interacting vacuum+matter system for the three cases: (I) $\alpha \neq 0$, $\beta = 0$, (II) $\alpha = 0$, $\beta \neq 0$, and (III) both $\alpha$ and $\beta$ are not zero.}
\label{tab:2-fluid}
\renewcommand\arraystretch{2.5}
\begin{tabular}{|@{~~}c@{~~}|c|@{~~}c@{~~}|@{~~}c@{~~}|@{~~}c@{~~}|@{~~}c@{~~}|}
\hline
\textbf{Case} & \textbf{Point}
& $x_p$                                               & $y_p$                                                                & $\lambda$             & \multicolumn{1}{c@{~~}|}{\textbf{Existence conditions}} \\ \hline
\multirow{2}{*}{$\alpha\neq0$, $\beta=0$}    & \textbf{Ia}   & $0$                                                 & $1$                                                                  & $-\frac{3+\alpha}{2}$ & $\forall\,\alpha$         \\ 
\cline{2-6} 
                                             & \textbf{Ib}   & $\sqrt{1+\frac{\alpha}{3}}$                         & $\epsilon\sqrt{-\epsilon\frac{\alpha}{3}}$                           & $3+\alpha$            & $\alpha >-3$                                       \\ \hline
\multirow{2}{*}{$\alpha=0$, $\beta\neq0$}    & \textbf{IIa}  & $1$                                                 & $0$                                                                  & $\frac{3+\beta}{2}$   & $\forall\,\beta$           \\ 
\cline{2-6} 
                                             & \textbf{IIb}  & $\sqrt{-\frac{\beta}{3}}$                           & $\epsilon\sqrt{\epsilon\left(1+\frac{\beta}{3}\right)}$   & $-(3+\beta)$          & $\beta < 0$                                    
\\ \hline
\multirow{2}{*}{$\alpha\neq0$, $\beta\neq0$} & 
	\textbf{III+} & 
	$\sqrt{\frac{(\alpha-\beta+3)+S(\alpha,\beta)}{6}}$ & 
	$\epsilon\sqrt{\frac{(-\alpha+\beta+3)-S(\alpha,\beta)}{6\epsilon}}$ & $S(\alpha,\beta)$     & 
\begin{tabular}[c]{@{}c@{}} $S^2(\alpha,\beta)\geq 0$, 
and \\[-5pt]
either $\beta>0$ or \\[-5pt]
$\beta<0$ with $\alpha-\beta+3 > 0$ 
\end{tabular}
\\ \cline{2-6} 
	& \textbf{III-} & 
    $\sqrt{\frac{(\alpha-\beta+3)-S(\alpha,\beta)}{6}}$ & 
    $\epsilon\sqrt{\frac{(-\alpha+\beta+3)+S(\alpha,\beta)}{6\epsilon}}$ & $-S(\alpha,\beta)$     & 
\begin{tabular}[c]{@{}c@{}} $S^2(\alpha,\beta)\geq 0$, 
and \\[-5pt]
$\alpha-\beta+3 > 0$ and $\beta<0$ 
\end{tabular}
 \\ \hline
\end{tabular}%
\end{table}


\section{Three-fluid system}
\label{sec:ref3fluid}

We will now consider a three-fluid cosmology with a noninteracting barotropic fluid with density $\rho_\gamma\geq0$, and pressure $P_\gamma=w_\gamma\rho_\gamma$, in addition to an interacting vacuum plus pressureless dark matter.
This would allow us to include, for example, the effect of radiation in the standard hot big bang cosmology as a barotropic fluid with $w_\gamma=1/3$, and we use this value to illustrate the qualitative evolution in our figures. In a realistic cosmology we require radiation corresponding to relativistic particles in the standard model to dominate the energy density of the early universe, and in particular at the time of primordial nucleosynthesis. However in our analysis we leave $w_\gamma$ undetermined.

The conservation equation for the barotropic fluid is
\begin{equation}
	\dot{\rho}_\gamma + 3H\rho_\gamma(1+w_\gamma) = 0
\end{equation}
with constant equation of state $w_\gamma$, while the Hubble expansion rate is now determined by the total energy density of all three fluids, so the Friedmann equation becomes
\begin{equation}
H^2 = \frac{\kappa^2}{3}\qty(\rho_\textrm{m} + V + \rho_\gamma ).
\label{Eqn:Friedmann-three-fluid}
\end{equation}

If we introduce a dimensionless density parameter for the barotropic fluid
\begin{equation}
	z \equiv \frac{\kappa\sqrt{\rho_\gamma}}{\sqrt{3}H}.
\end{equation}
then in terms of our dimensionless variables the Friedmann constraint Eq.~\eqref{Eqn:Friedmann-three-fluid} becomes
\begin{equation}
	x^2 + \epsilon y^2 = 1 - z^2.
	\label{Eqn:Constraint3fluid}
\end{equation}
Requiring $\rho_\gamma\geq0$ and hence $z^2\geq0$, our solutions lie within the unit circle $x^2+y^2\leq1$ for positive vacuum energy ($y>0$), or within the unit hyperbola $x^2-y^2\leq1$ for negative vacuum energy ($y<0$).


\subsection{Evolution equations and fixed points}

The evolution equations for three-fluid system are given by Eqs. \eqref{Eqn:Xprime-wq} and \eqref{Eqn:Yprime-wq} with an additional evolution equation for $z$
\begin{align}
    z' &= \frac{3}{2}\left( w-w_\gamma \right)z\,,
    \label{Eqn:Z3-fluidq} 
\end{align}
subject to the constraint~\eqref{Eqn:Constraint3fluid},
where the overall equation of state for three-fluid system is now given by
\begin{equation}
\label{Eqn:wthree}
	w = \frac{-V+w_\gamma\rho_\gamma}{\rho_\mathrm{m}+V+\rho_\gamma} 
	= w_\gamma(1-x^2)-(1+w_\gamma)\epsilon y^2.
\end{equation}
Equation \eqref{Eqn:Z3-fluidq} yields two conditions for the existence of fixed points, $z'\rightarrow 0$, in the three-fluid system, which are either
\begin{equation}
	z \rightarrow z_p = 0 \quad\text{or}\quad w \rightarrow w_p = w_\gamma.
\end{equation}

The first condition, $z_p=0$, requires that the barotropic fluid density become negligible and the three-fluid constraint \eqref{Eqn:Constraint3fluid} reduces to the two-fluid constraint \eqref{Eqn:ConstraintXY2fluid}. The two-fluid system is thus an invariant one-dimensional subspace of the two-dimensional three-fluid phase space. In particular the three-fluid fixed points remain as fixed points in the three-fluid system, and the eigenvalues in the two-fluid system remain eigenvalues in the full three-fluid phase space.
The overall equation of state is given by the two-fluid equation of state~\eqref{Eqn:wx}.

The second condition, $w_p=w_\gamma$, may yield additional fixed points which can be found from the existence condition \eqref{Eqn:qnovel}
\begin{equation}
\label{novel3fluidq}
q = 3w_\gamma x_p^2 = - 3(1+w_\gamma)\epsilon y_p^2  \,.
\end{equation}
In general this depends on the form of the dimensionless interaction $q$.
But in any interaction model for which $q\to0$ as $x\to0$ and $y\to0$ we have a fixed point where the barotropic fluid density dominates, $z\to1$, and hence
\begin{equation}
\label{def:fpO}
\mathbf{O}: \quad\quad x_{\mathrm{O}}=0\,,\quad  y_{\mathrm{O}}=0\,, \quad z_{\mathrm{O}}=1 \,.
\end{equation}
In particular this barotropic-fluid-dominated fixed point exists for all values of the parameters $\alpha$ and $\beta$ in our simple interaction model \eqref{Eqn:qsimple}, and it is the only fixed point that does not lie on the invariant (two-fluid) subspace $x^2+\epsilon y^2=1$.

On the other hand, a novel three-fluid fixed point may exist for other interaction models. For example, if we consider an alternative interaction model
\begin{equation}
\label{exampleq}
q = C x^2 y^2 \,,
\end{equation}
then there is a three-fluid scaling solution with $w_p=w_\gamma$ corresponding to the fixed point
\begin{equation}
x_p = \sqrt{\frac{3w_\gamma}{C}} \,, 
\quad 
y_p = \epsilon \sqrt{\frac{-3\epsilon(1+w_\gamma)}{C}} \,,
\end{equation}
which exists either for $0<C/3w_\gamma<1$ and $(1+w_\gamma)C>0$, or for $C/3w_\gamma>1$ and $0<-3(1+w_\gamma)/C<1$.

\subsection{Simple interaction model}

In this subsection we consider the simple interaction model \eqref{Eqn:qsimple} in the presence of a third barotropic fluid.
Using the constraint \eqref{Eqn:Constraint3fluid} and overall equation of state \eqref{Eqn:wthree} for the three-fluid system, the evolution equations, \eqref{Eqn:Xprime-wq} and \eqref{Eqn:Yprime-wq}, become
\begin{align}
	x' &= \left[3w_\gamma-\alpha -\left(3 (1+w_\gamma)+\frac{\beta}{x^2}\right)\epsilon y^2 - 3w_\gamma x^2\right]\frac{x}{2} 
	\label{Eqn:x-3fluid} \\
	%
	%
	y' &= \left[\beta + 3(1+w_\gamma)+
	\left(\epsilon\frac{\alpha}{y^2}-3w_\gamma\right)x^2
	-3(1+w_\gamma)\epsilon y^2 \right]\frac{y}{2}
	\label{Eqn:y-3fluid}
\end{align}
As remarked in the preceding subsection, the fixed points for this three-fluid system correspond to those of the two-fluid subspace ($z=0$) plus the barotropic-fluid-dominated ($z_{\mathrm{O}}=1$) fixed point \eqref{def:fpO}.
These fixed points are listed in Table \ref{tab:3-fluid}.

\begin{table}[]
\caption{Summarized table for fixed points, eigenvalues and their existence conditions in the three-fluid system.}
\resizebox{\textwidth}{!}{%
\begin{tabular}{|@{~~}c@{~~}|c|c|c|c|c|}
\hline
  \textbf{Case} & \textbf{Point} &
  $x_p$ &
  $y_p$ &
  $\lambda$ &
  \textbf{Existence conditions} \\ \hline
\multirow{2}{*}{\begin{tabular}[c]{@{}c@{}}$\alpha\neq0$, $\beta=0$\end{tabular}} &
  \textbf{Ia} &
  $0$ &
  $1$ &
  $-3(1+w_\gamma)$, \quad $-\frac{3+\alpha}{2}$ &
  $\forall\,\alpha$ and $\forall\,w_\gamma$ \\ \cline{2-6} 
 &
  \textbf{Ib} &
  $\sqrt{1+\frac{\alpha}{3}}$ &
  $\epsilon\sqrt{-\epsilon\frac{\alpha}{3}}$ &
  $\alpha - 3w_\gamma$, \quad $3+\alpha$ &
  $\alpha > -3$ and $\forall\, w_\gamma$ \\ \hline
\multirow{2}{*}{\begin{tabular}[c]{@{}c@{}}$\alpha=0$, $\beta\neq0$\end{tabular}} &
  \textbf{IIa} &
  $1$ &
  $0$ &
  $-3w_\gamma$, \quad $\frac{3+\beta}{2}$ &
  $\forall\,\beta$ and $\forall\,w_\gamma$ \\ \cline{2-6} 
 &
  \textbf{IIb} &
  $\sqrt{-\frac{\beta}{3}}$ &
  $\epsilon\sqrt{\epsilon\qty(1+\frac{\beta}{3})}$ &
  $-(3+\beta)-3w_\gamma$, \quad $-(3+\beta)$ &
  $\beta < 0$ and $\forall\, w_\gamma$ \\ \hline
\multirow{2}{*}{\begin{tabular}[c]{@{}c@{}}$\alpha\neq0$, $\beta\neq0$\end{tabular}} &
  \textbf{III+} &
  $\sqrt{\frac{(\alpha-\beta+3)+S(\alpha,\beta)}{6}}$ &
  $\epsilon\sqrt{\frac{(\beta-\alpha+3)-S(\alpha,\beta)}{6\epsilon}}$ &
  \begin{tabular}[c]{@{}c@{}}$-\frac{1}{2}\qty[\beta-\alpha+3+6w_\gamma - S(\alpha,\beta)]$, \\ $S(\alpha,\beta)$\end{tabular} &
  \begin{tabular}[c]{@{}c@{}}$S^2(\alpha,\beta)\geq 0$, and \\ either $\beta>0$ or \\ $\beta<0$ with $\alpha-\beta+3>0$\end{tabular} \\ \cline{2-6} 
 &
  \textbf{III-} &
  $\sqrt{\frac{(\alpha-\beta+3)-S(\alpha,\beta)}{6}}$ &
  $\epsilon\sqrt{\frac{(\beta-\alpha+3)+S(\alpha,\beta)}{6\epsilon}}$ &
  \begin{tabular}[c]{@{}c@{}}$-\frac{1}{2}\qty[\beta-\alpha+3+6w_\gamma + S(\alpha,\beta)]$, \\ $-S(\alpha,\beta)$\end{tabular} &
  \begin{tabular}[c]{@{}c@{}}$S^2(\alpha,\beta)\geq 0$, and \\ $\alpha-\beta+3 > 0$ and $\beta<0$\end{tabular} \\ \hline
\multicolumn{1}{|c|}{\begin{tabular}[c]{@{}c@{}}Both $\alpha\neq0$, $\beta=0$ \\ and $\alpha=0$, $\beta\neq0$, \\ and also $\alpha\neq0$, $\beta\neq0$\end{tabular}} & \textbf{O} &
  $0$ &
  $0$ &
  $\frac{1}{4}\qty[\beta - \alpha + 3+ 6w_\gamma \pm S(\alpha,\beta)]$ &
  $S^2(\alpha,\beta)\geq0$, and $\forall\,w_\gamma$ \\ \hline
\end{tabular}%
}
\label{tab:3-fluid}
\end{table}

We analyze the existence conditions for the fixed points and their stability for the same three cases as in the two-fluid system: (I) $\alpha\neq0$, $\beta=0$, (II) $\alpha=0$, $\beta\neq0$ and (III) $\alpha,\beta\neq0$. 
The fixed point \textbf{O} exists for all cases, while the existence conditions for the fixed points in the two-fluid subspace ($z_p=0$) remain unchanged. 
However, the stability of these fixed points in the three-fluid case now also depends on the barotropic equation of state, $w_\gamma$, as well as the values of $\alpha$ and $\beta$. 

\begin{figure*}
\begin{center}
\includegraphics[width=\textwidth]{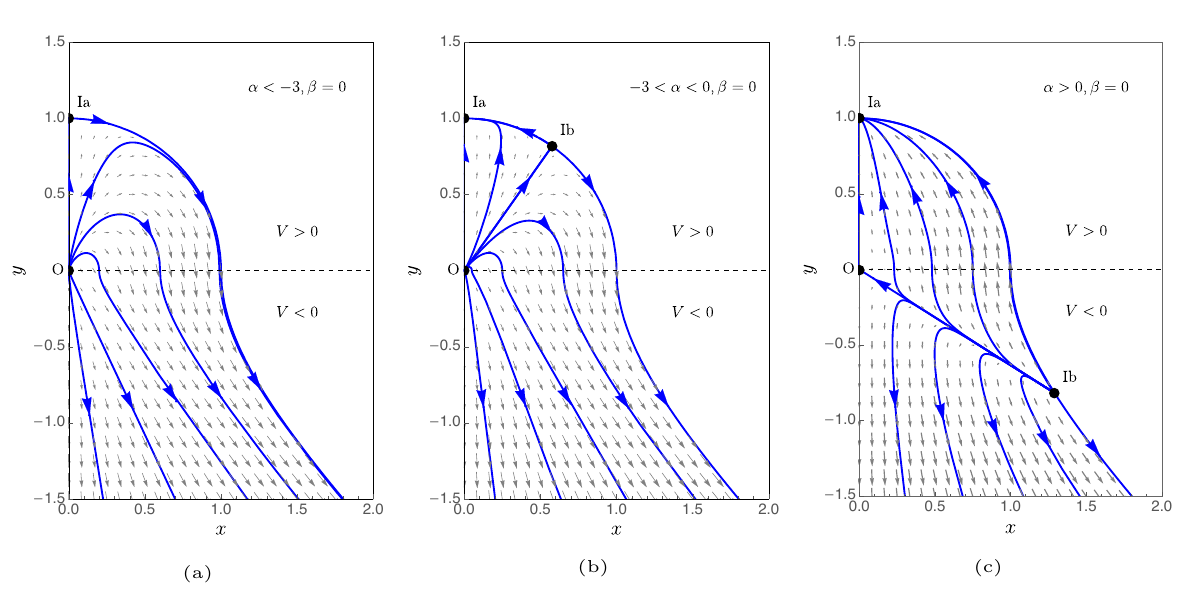}
\caption{The phase-space evolution of $(x,y)$ with respect to $N$ for the three-fluid system in case I ($\beta=0$) with radiation fluid, $w_\gamma=1/3$, illustrating the existence and stability of the fixed points in each range of $\alpha$.\label{Fig:3fluidcaseI}}
\end{center}
\end{figure*}

\begin{figure*}
\begin{center}
\includegraphics[width=\textwidth]{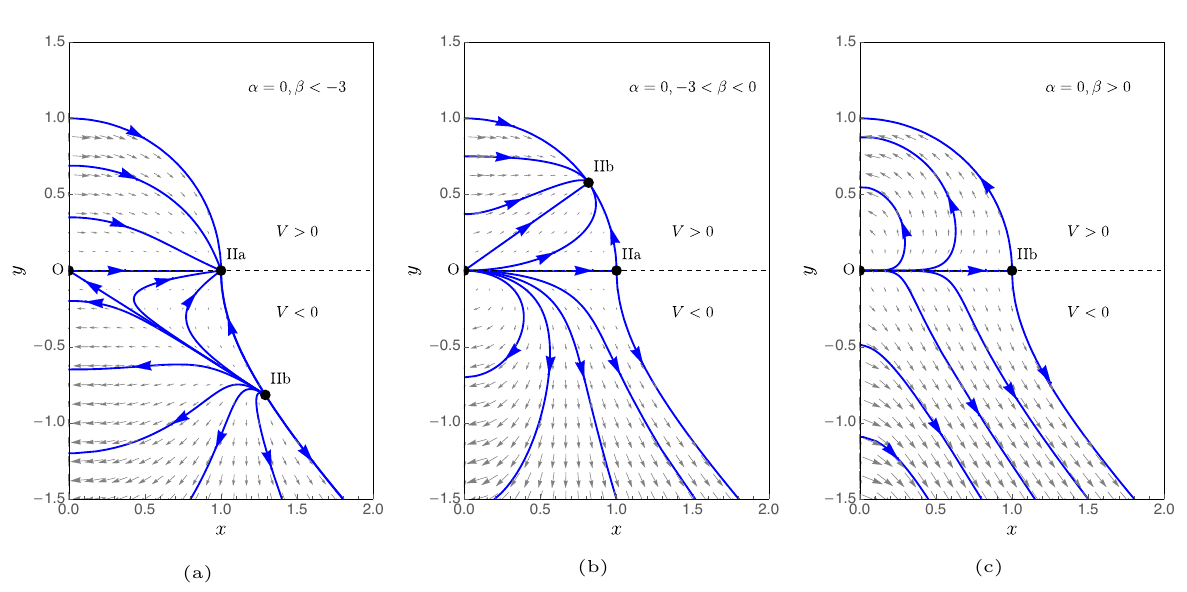}
\caption{The phase-space evolution of $(x,y)$ with respect to $N$ for the three-fluid system in case II ($\alpha=0$) with radiation fluid, $w_\gamma=1/3$, illustrating the existence and stability of the fixed points in each range of $\alpha$.\label{Fig:3fluidcaseII}}
\end{center}
\end{figure*}

\begin{figure*}
\begin{center}
\includegraphics[width=\textwidth]{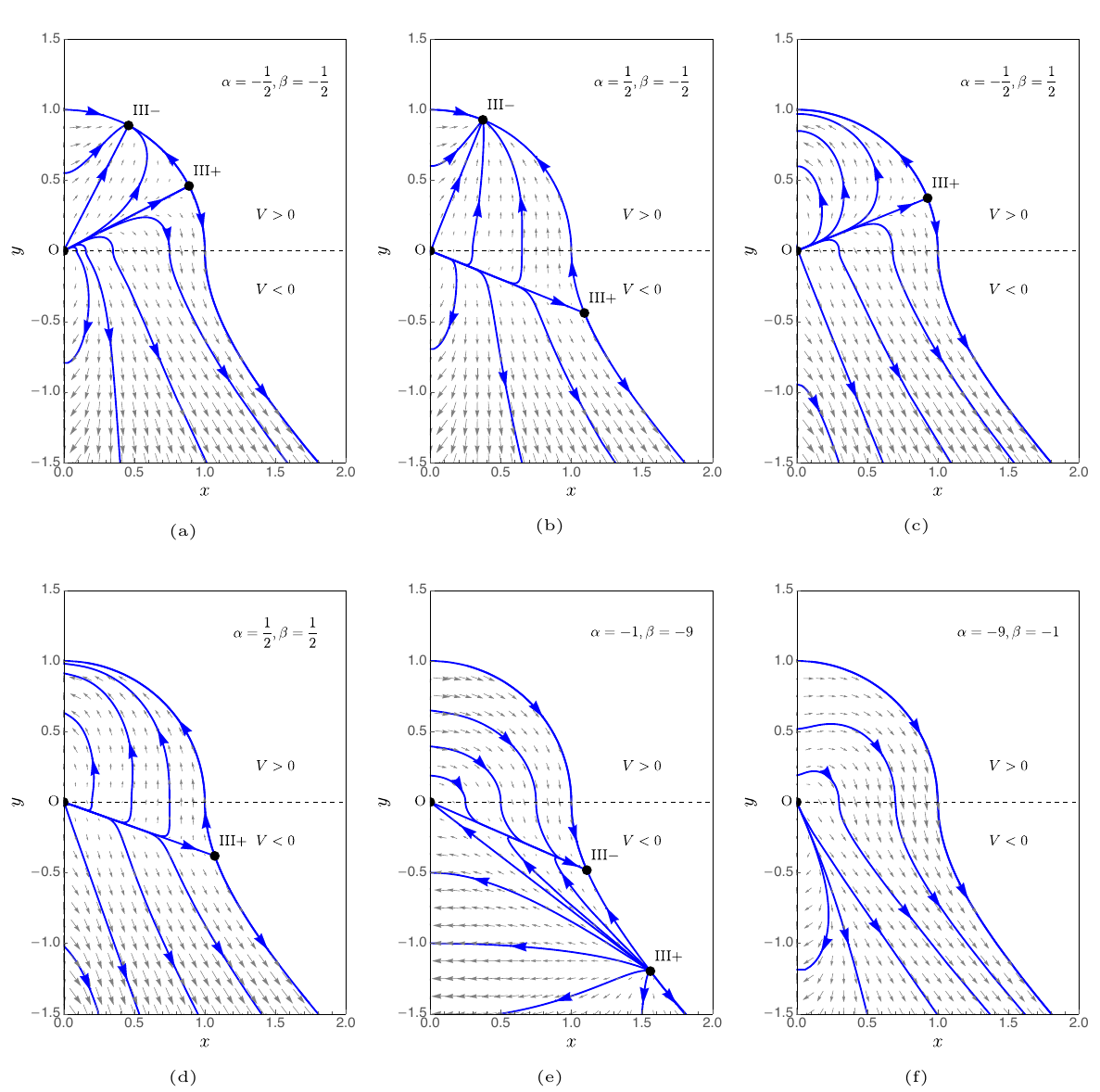}
\caption{\label{fig:gen-ab-rad} The phase-space evolution of $(x,y)$ with respect to $N$ for the three-fluid system in case III ($\alpha\neq0$, $\beta\neq0$) with radiation fluid, $w_\gamma=1/3$, illustrating the existence and stability of the fixed points. Each plot shows the phase space for different values of $\alpha$ and $\beta$ picked from each of the regions labelled a)-f) in Fig.~\ref{Fig:gen.case-condition}.}
\end{center}
\end{figure*}


In a two-dimensional phase space, $(x,y)$, the eigenvalues are usually calculated from the Jacobian matrix of the system of linearized evolution equations for small perturbations about each fixed point (see Appendix~\ref{appA}).
Considering small perturbations, $x\rightarrow x_p+u$ and $y\rightarrow y_p+v$, about the two-fluid fixed points with $z_p=0$, the linearized evolution Eqs. \eqref{Eqn:x-3fluid} and \eqref{Eqn:y-3fluid} for $u$ and $v$ can be written in matrix form
\begin{equation}
\renewcommand{\arraystretch}{1.2}
\mqty( u' \\ v' ) = J_p \mqty( u \\ v )
\end{equation}
where the Jacobian matrix
\begin{equation}
J_p(x_p,y_p) = 
\mqty[
-\frac{3+\alpha}{2}+\epsilon \frac{\beta}{2} \frac{y_{p}^{2}}{x_{p}^{2}}+\frac{3}{2}(1-2w_{\gamma}) x_{p}^{2} &
-\qty(\frac{\beta}{x_{p}}+3(1+w_{\gamma}) x_{p} ) \epsilon y_{p} \\
\left(\epsilon\frac{\alpha}{y_{p}}-3 w_{\gamma} y_{p}\right)x_p &
\frac{\beta}{2}+\frac{1}{2}\qty(3-\epsilon\frac{\alpha}{y_{p}^{2}} )x_{p}^{2}-3(1+w_{\gamma})\epsilon y_{p}^{2}
]
\label{Jacobian}
\end{equation}
and we have used the two-fluid fixed-point relation, $x_p^2+\epsilon y_p^2=1$ in the above equations.
In order to determine the stabilities of the fixed points, we evaluate the eigenvalues of the Jacobian matrix for each fixed point (see Appendix~\ref{Jab-matrices}).


However we will analyze separately the stability of the barotropic-fluid-dominated fixed point \textbf{O} 
since the evolution Eqs. \eqref{Eqn:x-3fluid} and \eqref{Eqn:y-3fluid} (and hence the corresponding Jacobian matrix) become ill defined as $x\to0$ and $y\to0$ simultaneously.
%
Setting $x=u$ and $y=v$ and letting $u\to0$ and $v\to0$ then the evolution Eqs. \eqref{Eqn:x-3fluid} and \eqref{Eqn:y-3fluid} give
\begin{align}
	 u' &\approx \frac{1}{2}\qty[3w_\gamma-\alpha - 
	\epsilon\beta\qty(\frac{v}{u})^2]u \,,
	\label{uprm}
	\\
	v' &\approx \frac{1}{2}\qty[3(1+w_\gamma)+\beta+\epsilon\alpha \qty(\frac{u}{v})^{2}] v \,.
	\label{vprm}
\end{align}
%
%
We keep the terms $u^2/v^2$ and $v^2/u^2$ since Eqs. \eqref{uprm} and \eqref{vprm} are undetermined in general when $u$ and $v$ both approach zero. 

We will consider trajectories which approach the fixed point, $x=u\to0$ and $y=v\to0$, at a fixed angle $\theta$ such that $t\equiv\tan\theta=v/u=$const and is finite.
This is only consistent if trajectories approach \textbf{O} along an eigenvector at angle $\theta$ with eigenvalue $\lambda_t$ such that $u'=\lambda_t u$ and $v'=\lambda_t v$, where the consistency of the evolution Eqs. \eqref{uprm} and \eqref{vprm} requires
\begin{equation}
	\lambda_t
	= \frac{1}{2}\qty[3w_\gamma-\alpha -\epsilon\beta t^2]
	= \frac{1}{2}\qty[3(1+w_\gamma)+\beta+\epsilon\alpha t^{-2}] \,.
	\label{eigenlt}
\end{equation}
Solving for $t^2$ yields
\begin{equation}
	t^2 = \frac{1}{2\epsilon\beta}
	\qty[-(\alpha +\beta +3) \mp S(\alpha,\beta)]
\end{equation}
where $S(\alpha,\beta)$ is defined in Eq. \eqref{Eqn:definedS}.
Substituting the $t^2$ equation back into Eqs. \eqref{eigenlt}, we find that the eigenvalues are given by
\begin{equation}
	\lambda_{\mathrm{O}\pm} 
	= \frac{1}{4}\qty[\beta - \alpha + 3 
	+ 6w_\gamma \mp S(\alpha,\beta) ] 
	\,.
	\label{eq:lambda-t}
\end{equation}
Thus we see that the fixed point \textbf{O} only has real eigenvalues for $S^2(\alpha,\beta)>0$. If $S(\alpha,\beta)$ is real then at least one of the fixed points \textbf{III$\pm$} exists and we have 
 \begin{equation}
	\lambda_{\mathrm{O}\pm} = \frac{3}{2}\qty(w_\gamma - w_{\mathrm{III}\pm})
\,.
\end{equation}


\subsubsection*{Case I: $\alpha\neq0,\beta=0$}

The phase space trajectories in this case are shown in Fig. \ref{Fig:3fluidcaseI}. For this case, the Jacobian matrix \eqref{Jacobian} becomes
\begin{equation}
J_{p,\mathrm{I}}
=\mqty[
-\frac{3+\alpha}{2} + \frac{3}{2}(1-2w_{\gamma})x_p^2  & 
-3\epsilon (1+w_{\gamma}) x_py_p   \\
 \epsilon\frac{\alpha x_p}{y_p}-3w_{\gamma} x_p y_p  & 
-3\epsilon (1+w_{\gamma})y_p^2 -\epsilon\frac{\alpha x_p^2}{2 y_p^2}+\frac{3}{2}x_p^2
].
\end{equation}

The vacuum-dominated ($y_{\mathrm{Ia}}=1$) fixed point \textbf{Ia} exists for any $\alpha$ and $w_\gamma$. 
It has eigenvalues
\begin{equation}
\lambda_{\mathrm{Ia},1} = -3(1+w_{\gamma}), \quad \lambda_{\mathrm{Ia},2} = -\frac{3+\alpha}{2}.
\label{eq:lambda-Ia12}
\end{equation}
Thus it is a
stable node for $\alpha>-3$ and $w_\gamma>-1$, 
an unstable node for $\alpha<-3$ and $w_\gamma<-1$ and
a saddle point for either $\alpha<-3$ and $w_\gamma>-1$, or for $\alpha>-3$ and $w_\gamma<-1$.

The interacting matter+vacuum scaling solution \textbf{Ib} exists for $\alpha>-3$ and any $w_\gamma$.
It has eigenvalues
\begin{equation}
\lambda_{\mathrm{Ib},1} = \alpha -3 w_{\gamma }, \quad \lambda_{\mathrm{Ib},2} = 3+\alpha
\end{equation}
Thus it is an
unstable node for $w_\gamma<\alpha/3$ and 
a saddle point for $w_\gamma>\alpha/3$.

The barotropic-fluid-dominated ($z_{\mathrm{O}}=1$) fixed point \textbf{O} exists for any $\alpha$ and $w_\gamma$. 
It has eigenvalues
\begin{equation}
\lambda_{\mathrm{O},1} = \frac32(1+w_\gamma) \,, 
\quad 
\lambda_{\mathrm{O},2} = \frac12 (3 w_\gamma -\alpha)
	\,.
	\label{eq:lambda-tI}
\end{equation}
Thus it is a 
stable node for $\alpha>3w_\gamma$ and $w_\gamma<-1$,
an unstable node for $\alpha<3w_\gamma$ and $w_\gamma>-1$, and 
it is a saddle point for either $\alpha>3w_\gamma$ and $w_\gamma>-1$, or for 
	$\alpha<3w_\gamma$ and $w_\gamma<-1$.

\subsubsection*{Case II: $\alpha=0,\beta\neq0$}

The phase space trajectories in this case are shown in Fig. \ref{Fig:3fluidcaseII}. For $\alpha=0,\beta\neq0$, the Jacobian matrix \eqref{Jacobian} reduces to
\begin{equation}
J_{p,\mathrm{II}} = 
\mqty[
-\frac{3}{2}+\epsilon \frac{\beta}{2} \frac{y_{p}^{2}}{x_{p}^{2}}+\frac{3}{2}(1-2w_{\gamma}) x_{p}^{2} ~~ &
-\qty(\frac{\beta}{x_{p}}+3(1+w_{\gamma}) x_{p} ) \epsilon y_{p} \\
-3 w_{\gamma}x_p y_{p} &
\frac{\beta}{2}+\frac{3}{2}x_{p}^{2}-3(1+w_{\gamma})\epsilon y_{p}^{2},
]
\,.
\end{equation}

The matter-dominated ($x_{\mathrm{IIa}}$) fixed point \textbf{IIa} exists for $\beta$ and $w_\gamma$. It has  eigenvalues
\begin{equation}
\lambda_{\mathrm{IIa},1} = -3 w_{\gamma }, \quad \lambda_{\mathrm{IIa},2} = \frac{3+\beta}{2}
\label{eq:lambda-IIa12}
\end{equation}
Thus it is a stable node for $\beta<-3$ and $w_\gamma>0$. It is an unstable node for $\beta>-3$ and $w_\gamma<0$, and a saddle point for either $\beta>-3$ and $w_\gamma>0$ or for $\beta<-3$ and $w_\gamma<0$.

The interacting matter+vacuum scaling solution \textbf{IIb} exists for $\beta<0$ and any $w_\gamma$.
It has eigenvalues
\begin{equation}
\lambda_{\mathrm{IIb},1} = -\beta -3(1+w_{\gamma }), 
\quad
\lambda_{\mathrm{IIb},2} = -(3+\beta).
\label{eq:lambda-IIb}
\end{equation}
Thus it is a stable node for $\max\{-3(1+w_\gamma),-3\}<\beta<0$, an unstable node for $\beta<\min\{-3(1+w_\gamma),-3\}$, and a saddle point for either $-3(1+w_\gamma)<\beta<-3$ if $w_\gamma>0$ or for $-3<\beta<-3(1+w_\gamma)$ if $w_\gamma<0$.

The barotropic-fluid-dominated ($z_{\mathrm{O}}=1$) fixed point \textbf{O} exists for any $\alpha$ and $w_\gamma$. 
It has eigenvalues
\begin{equation}
\lambda_{\mathrm{O},1} = \frac32 w_\gamma \,, 
\quad 
\lambda_{\mathrm{O},2} = \frac12 (3 w_\gamma + 3 + \beta)
	\,.
	\label{eq:lambda-tII}
\end{equation}
Thus it is a 
stable node for $w_\gamma<\min\{0,-1-(\beta/3)\}$, an unstable node for $w_\gamma>\max\{0,-1-(\beta/3)\}$ and 
a saddle point for either $-1-(\beta/3)<w_\gamma<0$ if $\beta>-3$ or for $0<w_\gamma<-1-(\beta/3)$ if $\beta<-3$.

\subsubsection*{Case III: $\alpha,\beta\neq0$}

As in the two-fluid model, the interacting matter+vacuum scaling solutions \textbf{III$\pm$} for $\alpha\neq0$ and $\beta\neq0$ exist only if $S(\alpha,\beta)$ defined in Eq.~\eqref{Eqn:definedS} is real and $x_{\mathrm{III}\pm}$ defined in Eq. \eqref{xIII:defined} is real [see Eq.~\eqref{Eqn:xpcondition}].

The Jacobian matrix \eqref{Jacobian} does not immediately simplify.
%
The characteristic equation for the eigenvalues at each point can be written in the standard form
\begin{equation}
\lambda^2 - \tr(J_p)\lambda + \det(J_p) = 0,
\label{charateristiclambda}
\end{equation}
where the trace and determinant of the Jacobian matrix \eqref{Jpelements} at each point \textbf{III$\pm$} are given by
\begin{align}
\tr(J_p) &= \pm\frac{3}{2}S(\alpha,\beta)-\frac{1}{2}
(\beta-\alpha +3 + 6w_{\gamma}) 
\,,\\
\det(J_p) &= \frac{1}{2}S^2(\alpha,\beta) \mp \frac{1}{2}(\beta-\alpha+3+6w_\gamma) S(\alpha,\beta) \,.
\end{align}
%

%
%
%

Thus we find the two eigenvalues at each point are given by
%
\begin{equation}
\lambda_{\mathrm{III}\pm,1} = \pm S(\alpha,\beta) 
\,,
\quad
\lambda_{\mathrm{III}\pm,2} 
= -\frac{1}{2}\qty[\beta-\alpha+3+6w_\gamma \mp S(\alpha,\beta)]
\,.
\end{equation}
The equation of state at the fixed points \textbf{III$\pm$} is given by Eq. \eqref{wIII:defined}, and hence we have 
\begin{equation}
\lambda_{\mathrm{III}\pm,2} = -3(w_\gamma-w_{\mathrm{III}\pm}) \,.
\end{equation}

The point \textbf{III+} is always unstable, when it exists, since $S(\alpha,\beta)>0$. It is a saddle point if $w_{\mathrm{III}+}<w_\gamma$, and otherwise it is an unstable node.
The point \textbf{III-} is a stable node if $w_{\mathrm{III}-}<w_\gamma$, and otherwise it is a saddle point.

The barotropic-fluid-dominated ($z_{\mathrm{O}}=1$) fixed point \textbf{O} exists for any $\alpha$, $\beta$ and $w_\gamma$. 
It has eigenvalues
\begin{equation}
\lambda_{\mathrm{O},\pm} = \frac32 (w_{\mathrm{III}\pm}-w_\gamma) 
	\,.
	\label{eq:lambda-tIII}
\end{equation}
where $w_{\mathrm{III}\pm}$ is defined by Eq. \eqref{wIII:defined} even in cases where the fixed points \textbf{III$\pm$} may not exist.
If $S^2(\alpha,\beta)>0$ then $w_{\mathrm{III}+}<w_{\mathrm{III}-}$ and the fixed point \textbf{O} is a stable node for $w_\gamma>w_{\mathrm{III}-}$, a saddle point for $w_{\mathrm{III}+}<w_\gamma<w_{\mathrm{III}-}$ and an unstable node for $w_\gamma<w_{\mathrm{III}+}$.

\begin{figure}[!h]
\centering
\includegraphics{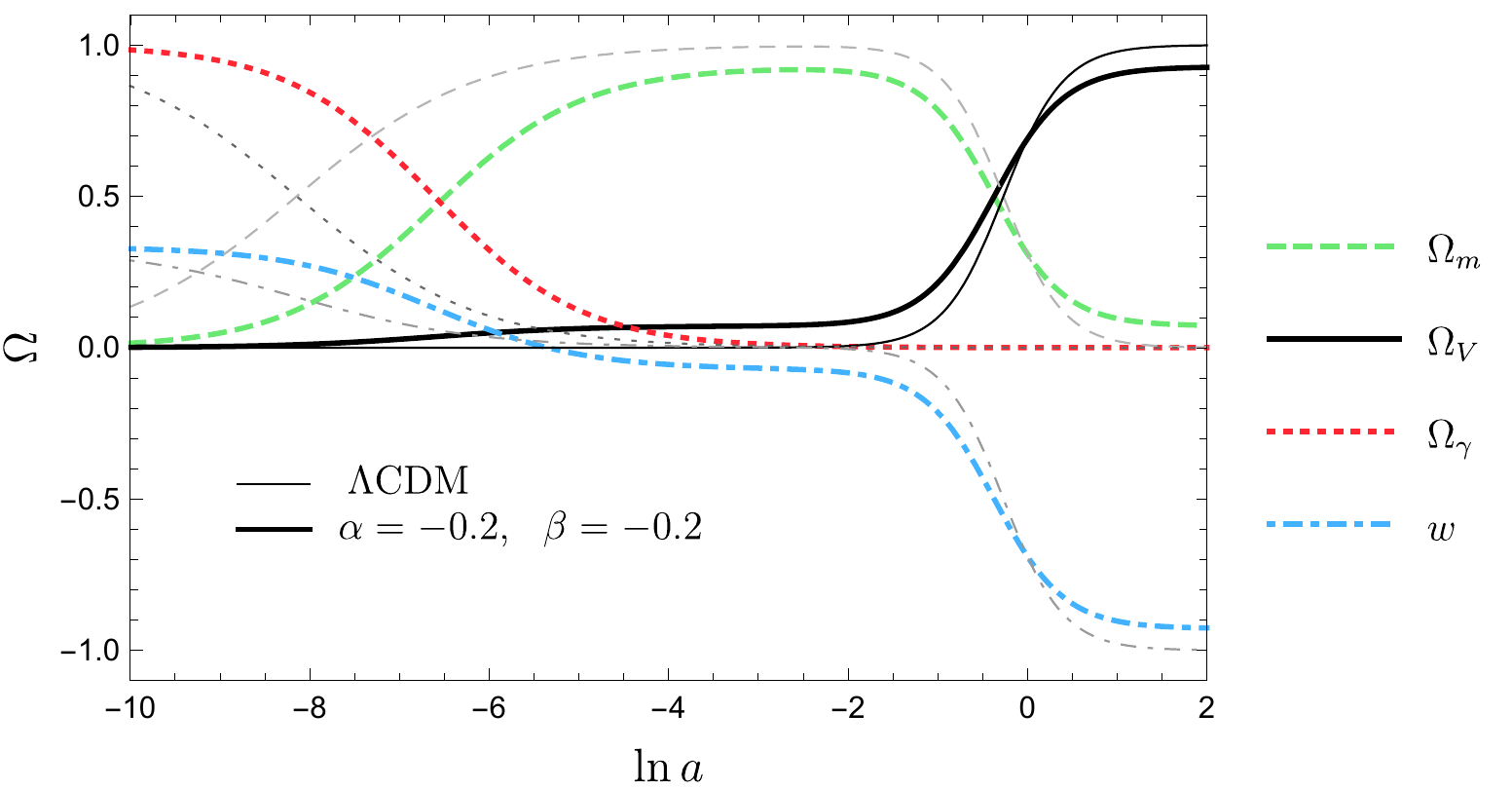}
\caption{\label{Omega-plot}
The evolution with respect to $N=\ln(a)$ of the dimensionless density parameters for vacuum energy (black, solid line), matter density (green, dashed line) and radiation (red, dotted line), in the three-fluid cosmology with the simple interaction model (\ref{modelQ}) with $\alpha=-0.2$ and $\beta=-0.2$.}
\end{figure}


%

\section{Discussion}
\label{sec:discuss}

In this work we have presented a phase-plane analysis of the cosmic evolution in a spatially flat FRW spacetime with an interacting vacuum energy scenario, allowing energy exchange between pressureless matter and vacuum energy. We have allowed for the vacuum energy to be either positive or negative, though late-time acceleration requires a positive vacuum energy. 
Using dimensionless variables $x$ and $y$ defined in Eq.~(\ref{Var:xy}), analogous to those used to study scalar field cosmologies \cite{Copeland:1997et}, fixed points in the phase space correspond to power-law scaling solutions dominated by one or more components. 
We were able to give existence conditions for the fixed points in general, without specifying the functional form of the interaction, $Q$. 
We analyzed both two-fluid (pressureless matter + vacuum energy) and three-fluid systems, including an additional noninteracting barotropic fluid (e.g., radiation). 

For two fluids, corresponding to the interacting matter and vacuum energies, the fixed-point condition (\ref{Eqn:q2fluidcondition}) requires that along the scaling solution we have
\begin{equation}
\left. \frac{Q}{H} \right|_p = \left.  -\kappa^2 \rho_\mathrm{m} V \right|_p \,.
\end{equation}
In the absence of an interaction ($Q=0$) this includes the trivial cases of matter-dominated ($V=0$) or a vacuum-dominated ($\rho_\mathrm{m}=0$) cosmologies. But it also allows for the existence of nontrivial scaling solutions with $V\neq0$ and $\rho_\mathrm{m}\neq0$ in interacting models where $Q=0$.

The two-fluid system is an invariant subspace corresponding to the one-dimensional boundary of the three-fluid system. Hence all the fixed points found in the two-fluid system are also fixed points in the three-fluid system. 
However in the presence of a third barotropic fluid with equation of state $P_\gamma=w_\gamma\rho_\gamma$ we can obtain additional fixed point solutions which obey the condition (\ref{novel3fluidq}), which we can write as
\begin{equation}
\left. \frac{Q}{H} \right|_p = \left.  3w_\gamma \rho_\mathrm{m} \right|_p=  \left.  - 3(1+w_\gamma)V \right|_p \,.
\end{equation}
Again there is one trivial case $\rho_\mathrm{m}=V=0$ corresponding to a barotropic-fluid-dominated fixed point, $(x=0,y=0,z=1)$, which exists for any barotropic index $w_\gamma$ in the noninteracting limit $Q/H\to0$.
However there exist nontrivial three-fluid scaling solutions in nonlinear interaction models such as that given in Eq.~(\ref{exampleq}) which corresponds to a model in which $Q\propto H \rho_\mathrm{m} V$.

To illustrate the full phase-space evolution. including the stability of the fixed points, we focused on a simple linear interaction model $Q= \alpha H\rho_\mathrm{m}+\beta HV$, identifying fixed points in each case when $\alpha\neq0$ and $\beta=0$, $\alpha=0$ and $\beta\neq0$, or when both $\alpha$ and $\beta$ are nonzero. 
For the interaction model $Q=\alpha H\rho_\mathrm{m}$, a vacuum-dominated solution (\textbf{Ia}) exists for which $y_\mathrm{Ia}=1$ and $w_\mathrm{Ia}=-1$. This is a late-lime attractor if $\alpha>-3$. 
%
%
%
Conversely, in the case $Q=\beta HV$, there exists a matter-dominated solution \textbf{IIa} for which $x_\mathrm{IIa}=1$ and $w_\mathrm{IIa}=0$. 
This is unstable for $\beta>-3$ in which case the universe can evolve towards a fixed point (for $-3<\beta<0$) with negative pressure (due to the presence of vacuum energy) at late times.
%
The only additional fixed point in the presence of a third noninteracting barotropic fluid given this interaction model is the barotropic-fluid-dominated fixed point. 

In the general case when $\alpha$ and $\beta$ are both nonzero, one possible cosmology corresponds to small negative values of the  parameters $\alpha$ and $\beta$ corresponding to the region (a) shown in Fig.~\ref{Fig:gen.case-condition}. In this case the generic early-time solution is radiation dominated (assuming a noninteracting barotropic fluid with $w_\gamma=1/3$).
The fixed points \textbf{III+} and \textbf{III-} shown in Fig.~\ref{fig:gen-ab-rad} are a saddle point and an attractor respectively in the two-fluid subspace system. In this case the overall equation of state at the fixed points is in the form
\begin{equation}
w_{\mathrm{III}\pm} = - \frac{1}{6} \qty[\beta-\alpha+3 \mp S(\alpha,\beta)].
\end{equation}
where $S(\alpha,\beta)$ is defined in Eq.~(\ref{Eqn:definedS}). For small $\alpha$ and $\beta$ we have $S(\alpha,\beta)\simeq 3+\alpha+\beta$ and hence
\begin{equation}
w_{\mathrm{III}+} \simeq \frac{\alpha}{3} \,, \qquad w_{\mathrm{III}-}  \simeq -1 - \frac{\beta}{3} \,.
\end{equation}
Thus we have a cosmological solution that evolves from radiation domination (\textbf{O}) to an intermediate matter-vacuum scaling solution (\textbf{III+}) dominated by the matter density for small $\alpha$ to a matter-vacuum scaling solution (\textbf{III-}) dominated by vacuum energy density at late times, which is accelerating for small $\beta$.
%
%
One such solution is shown in Fig.~\ref{Omega-plot} where we plot the dimensionless density parameters 
$\Omega_\mathrm{m}=x^2$, $\Omega_\mathrm{V}=y^2$ and $\Omega_\gamma=z^2$.

Our linear interaction model (\ref{modelQ}) is simple enough that we can integrate the coupled continuity equations~(\ref{Eqn:ContinuityVac}) and~(\ref{Eqn:ContinuityMat}) to obtain closed form solutions for the energy densities \cite{Quercellini:2008vh}, such that (for $S$ real) we have
\begin{eqnarray}
\rho_\mathrm{m} &=& \left[ \left(\frac{3+\alpha+\beta+S}{2S}\rho_{\mathrm{m},0}+\frac{\beta}{S}\rho_{\mathrm{V},0}\right)a^{-S/2}
- \left(\frac{3+\alpha+\beta-S}{2S}\rho_{\mathrm{m},0}+\frac{\beta}{S}\rho_{\mathrm{V},0}\right)a^{+S/2} \right] a^{-(3+\alpha-\beta)/2} \,,
\nonumber \\
\rho_\mathrm{V} &=& \left[  \left( \frac{\alpha}{S}\rho_{\mathrm{m},0} +\frac{3+\alpha+\beta+S}{2S}\rho_{\mathrm{V},0} \right)a^{+S/2} 
 - \left( \frac{\alpha}{S}\rho_{\mathrm{m},0} +\frac{3+\alpha+\beta-S}{2S}\rho_{\mathrm{V},0} \right)a^{-S/2} \right] a^{-(3+\alpha-\beta)/2} \,.
 \end{eqnarray}
We can then identify the fixed points in our phase space with the early- and late-time limits of this closed-form solution.
But for more general nonlinear interactions we have presented a qualitative approach that can be used to find scaling solutions and their stability without requiring a closed form solution. 

While there are many different theoretical models to describe the evolution of a late-time accelerating cosmology, these models are potentially distinguishable through the evolution of perturbations. In particular the formation of structure in the late Universe is sensitive to the dark matter-vacuum interaction, while the presence of an interaction in the early Universe could affect the epoch of matter-radiation equality and thus the pattern of anisotropies in the cosmic microwave background. Indeed such early dark energy models have been proposed as one possible mechanism to resolve the tension between locally measured values of the Hubble parameter and those inferred from the cosmic microwave background~\cite{Knox:2019rjx}. We plan to explore the behavior of inhomogeneous perturbations about generic phase space trajectories in interacting vacuum cosmologies in future work.

%


 

\section*{Acknowledgments}

The authors are grateful to Marco Bruni for useful discussions.
C. K. is funded by the Royal Thai Government. 
P. R. is supported by the 13th Royal Golden Jubilee Ph.D. (RGJ-Ph.D.) scholarship awarded by TRF.
B. G. is supported by a TRF Basic Research Grant no.~BRG6080003 (TRF Advanced Research Scholar) and the Royal Society-Newton Advanced Fellowship (NAF-R2-180874).
H. A., J. S. and D. W. are supported by the Science and Technology Facilities Council grant ST/S000550/1.

\appendix

\section{Phase-plane autonomous systems}
\label{appA}

Supposed we have an autonomous system given by
\begin{equation}
x' = f(x,y), \qquad y' = g(x,y).
\label{autonomousEq}
\end{equation}
To analyze the stability of the fixed points, we consider first-order perturbations $u$ and $v$ around such fixed points
$x \rightarrow x_p + u$, $y \rightarrow y_p+v$. The autonomous system becomes
\begin{align}
u' &\approx f(\mathbf{x}_p) + \eval{\pdv{f}{x}}_{\mathbf{x}_p} u
+ \eval{\pdv{f}{y}}_{\mathbf{x}_p} v \label{uprime} \\
v' &\approx g(\mathbf{x}_p) + \eval{\pdv{g}{x}}_{\mathbf{x}_p} u
+ \eval{\pdv{g}{y}}_{\mathbf{x}_p} v \label{vprime}
\end{align}
where $\mathbf{x}_p=\{x_p,y_p\}$. Substituting these perturbed quantities back into the system \eqref{autonomousEq} yields the linear differential equation written in the form
\begin{equation}
\renewcommand{\arraystretch}{1.2}
\begin{bmatrix}
u' \\ v'
\end{bmatrix}
= J_p
\begin{bmatrix}
u \\ v
\end{bmatrix}
\end{equation}
where $J_p \equiv J(\mathbf{x}_p)$ is a $2\times2$ Jacobian matrix at the fixed points written in the form
\begin{equation}
J_p = \begin{bmatrix}
		 \pdv{f}{x}  &  \pdv{f}{y}  \\
	 	 \pdv{g}{x}  &  \pdv{g}{y} 
	  \end{bmatrix}_{\mathbf{x}=\mathbf{x}_p}.   
\end{equation}
The eigenvalues of the matrix $J_p$ can be written as a characteristic equation as
\begin{equation}
\lambda^2 - \tr(J_p)\lambda + \det(J_p) = 0,
\end{equation}
giving two eigenvalues $\lambda_{\pm}$, and the general solutions $u$ and $v$ can be expressed as
\begin{align}
u &= u_{+}e^{\lambda_{+}N} + u_{-}e^{\lambda_{-}N} \\
v &= v_{+}e^{\lambda_{+}N} + v_{-}e^{\lambda_{-}N},
\end{align}
with the constants $u_{\pm}$ and $v_{\pm}$. Thus the stability requires that the real parts of $\lambda_\pm$ are both negative.

\section{Jacobian matrices for fixed points}
\label{Jab-matrices}

\noindent{Point: $\mathbf{Ia:}\ (x_p=0,y_p=1)$}
\begin{equation}
J_{Ia} = \mqty[ -\frac{3+\alpha}{2} & 0 \\
			  0 & -3(1+w_\gamma) ]
\end{equation}

\noindent{Point: $\mathbf{IIa:}\ (x_p=1,y_p=0)$}
\begin{equation}
J_{IIa} = \begin{bmatrix}
 -3 w_{\gamma } & 0 \\
 0 & \frac{3+\beta}{2} \\
\end{bmatrix}
\end{equation}

\noindent{Point: $\mathbf{Ib:}\ (x_p=\sqrt{1+\frac{\alpha}{3}}, y_p = \epsilon\sqrt{-\epsilon\frac{\alpha}{3}})$}
\begin{equation}
J_{Ib} =
\begin{bmatrix}
 -(3+\alpha)w_{\gamma} & 
 -(1+w_{\gamma})\sqrt{-\epsilon\alpha(3+\alpha) }  \\
 -\epsilon(1+w_{\gamma})\sqrt{-\epsilon\alpha(\alpha+3)} &  
 3+(2+w_{\gamma})\alpha 
\end{bmatrix}
\end{equation}

\noindent{Point: $\mathbf{IIb:}\ (x_p=\sqrt{-\frac{\beta}{3}}, y_p = \epsilon\sqrt{\epsilon(1+\frac{\beta}{3})}$}
\begin{equation}
J_{IIb} = \begin{bmatrix}
 \beta(w_{\gamma}-1)-3 & 
 -w_{\gamma}\sqrt{-\beta\epsilon(3+\beta)}  \\
 -\epsilon w_{\gamma}\sqrt{-\beta\epsilon(3+\beta)}  &
 -(1+w_{\gamma})(3+\beta)
\end{bmatrix}
\end{equation}
%

\noindent{Point: $\mathbf{III\pm:}\ (x_{\pm}=\sqrt{\frac{\alpha-\beta+3 \pm S(\alpha,\beta)}{6}}, y_{\pm} = \sqrt{\frac{\beta-\alpha+3 \mp S(\alpha,\beta)}{6\epsilon}})$.} 
\begin{equation}
\left.
\begin{aligned}
J_{\pm,11} &= -\frac{1}{2}\qty[\alpha+\beta+3 -(\alpha-\beta+3)w_{\gamma} \mp \frac{1}{2}(1-w_{\gamma })S(\alpha,\beta)] \\
J_{\pm,12} &= -\frac{1}{12}[-\alpha +\beta+3 +6 w_{\gamma}\pm S(\alpha,\beta)] \sqrt{\epsilon [\alpha -\beta-3 \pm S(\alpha,\beta)] [-\alpha +\beta-3 \mp S(\alpha,\beta)]} 
\\
J_{\pm,21} &= -\frac{1}{12}\epsilon[-\alpha +\beta+3 +6 w_{\gamma} \pm S(\alpha,\beta)] \sqrt{\epsilon [\alpha -\beta-3 \pm S(\alpha,\beta)] [-\alpha +\beta-3 \mp S(\alpha,\beta)]} \\
J_{\pm,22} &= \alpha +\frac{1}{2} (\alpha -\beta -3) w_{\gamma } \pm \frac{1}{2}(2+w_{\gamma})S(\alpha,\beta)
\end{aligned} \quad\right\}
\label{Jpelements}
\end{equation}

%



\bibliographystyle{unsrt}
\bibliography{ref-new}

\end{document}